\documentclass{article} 

\usepackage{graphicx}

\usepackage{url}

\usepackage{natbib}

\begin{document}

\title{PDBCirclePlot: A Novel Visualization Method for Protein Structures}

\author{Francis Bell, 
Chunyu Zhao, and 
Ahmet Sacan\footnote{to whom correspondence should be addressed} }

\date{Biomedical Engineering, Drexel University, 3141 Chestnut St, Philadelphia, PA, 19104, USA }
\maketitle
\begin{abstract}

Interactive molecular graphics applications facilitate analysis of three dimensional protein structures. Naturally, non-interactive 2-D snapshots of the protein structures do not convey the same level of geometric detail. Several 2-D visualization methods have been in use to summarize structural information, including contact maps and 2-D cartoon views. We present a new approach for 2-D visualization of protein structures where amino acid residues are displayed on a circle and spatially close residues are depicted by links. Furthermore, residue-specific properties, such as conservation, accessibility, temperature factor, can be displayed as plots on the same circular view.\\
\\PDBCirclePlot is available at
 
\url{http://sacan.biomed.drexel.edu/pdbcircleplot}
\\Contact: \url{ahmet.sacan@drexel.edu}
\end{abstract}

\section{Introduction}
Three-dimensional structures of proteins are experimentally determined at atomic-level resolution by X-ray chrystallography and NMR spectroscopy methods. The Protein Data Bank (PDB) \citep{berman00} is a public repository of experimentally determined three-dimensonal protein structures, providing an invaluable resource for structure-based functional and evolutionary analysis. Numerous molecular graphics programs have been developed for 3D visualization of these structures, including PyMol \citep{delano02} and Jmol \citep{hanson13}. Although these programs are useful for interactive visualization in 3D, their static two-dimensional projections do not convey the structural geometry with the same accuracy and clarity. As a result, alternative simplified 2D representations of protein structures are often used to depict them in web pages, reports, and presentations. These representations emphasize different properties.

Contact maps display the inter-residue distances as a matrix, highlighting residue pairs that are in close spatial proximity and aiding in identification of domains and secondary structure elements (SSE) \citep{levitt77}.
Linear block diagrams have been used to display structural domains and functional residues in proteins \citep{todd99}.
Topology diagrams display SSEs or domains in a (non-linear) 2D cartoon, preserving three dimensional proximities among these elements as much as possible \citep{laskowski09,michalopoulos04}, and are often used to depict structural fold families defined by structural classification systems \citep{murzin95}.

In this work, we introduce a new visualization method for protein structures based on the Circos \citep{krzywinski09} diagrams, which was originally been developed for aesthetic presentation of genomic data but has also found use in other application domains. We capture the geometric information in the form of links between spatially proximal residues, using a color scale to represent the Euclidean distance between residues. Furthermore, unlike other 2D representations, we additionally display different properties of the amino acid residues as plots on the same circular view.

\section{Methods and Results}
\begin{figure*}[!t]
\centering 
\includegraphics[width=6in]{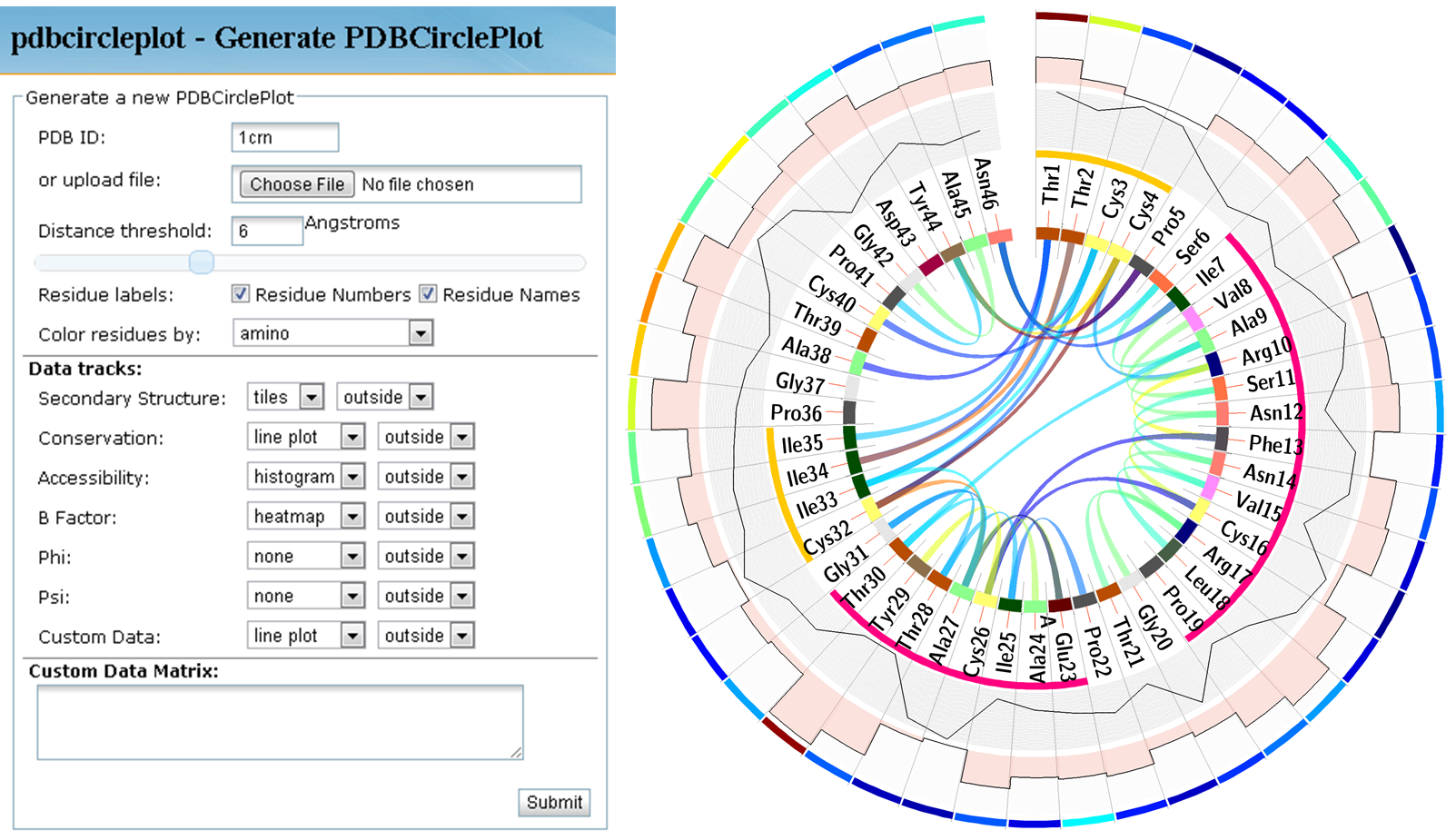}
\caption{PDBCirclePlot for PDB ID: 1CRN. The user can select how the residues are shown and labeled and which additional features are plotted around the protein. Pairs of residues closer to each other than the specified distance threshold are linked on the plot, with increasing Euclidean distance values shown by a color scale: beginning with dark blue and ranging through shades of blue, green, yellow, and red. The individual amino acids are colored by the amino acid type. Features associated with residues or custom user-provided data can be shown in data tracks using various plot types. These data tracks can be drawn inside or outside of the main circle representing the amino acid residues, following the same order shown on the web form. In this example, secondary structure, conservation, accessibility, and B Factor are shown outside; using tiles, line plot, histogram, and heatmap, respectively.}
\label{pdbcircleplotsnapshot}
\end{figure*}

PDBCirclePlot is provided as a web service, available from \url{http://sacan.biomed.drexel.edu/pdbcircleplot}. Figure~\ref{pdbcircleplotsnapshot} shows a snapshot of the web interface for specifying parameters and the output image generated. The user can upload a PDB-formated file or specify a PDB identifier with optional chain and range selection. For example, "1AAP A1-30 B10-40" will display the residues 1-30 from chain A and residues 10-40 from chain B of the protein 1AAP. 
The residues are shown along a circle, with each protein chain displayed as a separate ``ideogram''. Residues can be labeled by the amino acid names and/or residue numbers; and can be colored by amino acid type, chain id, sequence order, various residue properties, or using common coloring schemes used by 3D visualization methods, particularly those available in Jmol \citep{hanson13}.

A distance threshold on pairs of alpha-carbon atoms is used to determine which residues are linked in the circular view. The default threshold is 6\AA, which is a typically used threshold for defining residue contacts \citep{levitt77,heinig04}.
We have found 6\AA to highlight secondary structure elements and to capture other non-local spatial contacts. The links are colored using a blue-green-red color gradient, with close residue pairs linked in blue ribbons and far away residues (up to the distance threshold) shown in red. Links between consecutive residues, which are close to each other by the virtue of having a small separation along the backbone are not shown. Similar to contact maps, the secondary structure elements of the protein are apparent in the patterns of links. Alpha helices appear as an array of short consecutive links, whereas beta sheets appear as stretches of residues connected in parallel or anti-parallel.

An important difference from other 2D visualization methods is PDBCirclePlot's ability to display other structural or non-structural properties, integrated on the same view. The user can display secondary structure elements, accessibility, residue conservation, temperature factor, and Phi and Psi torsion angles. The user can also submit a custom data matrix to display other properties of interest. Each of these properties are shown using ``data tracks`` in Circos, displayed as line plots, histograms, or heatmaps.

The secondary structure elements are displayed using tiles, with alpha helices shown in pink and beta sheets shown in gold. The secondary structure definitions are extracted from the PDB header if available, otherwise calculated using DSSP \citep{kabsch83}. Solvent accessibility of residues are also calculated using DSSP. Residue conservation is obtained from the HSSP database \citep{joosten11} if available, or calculated from results of a PSI-BLAST search \citep{altschul97} against the NCBI's non-redundant protein database. Temperature factors are extracted and torsion angles are calculated directly from the PDB file.

PDB file handling and data collection for different residue properties are implemented in MATLAB. Location and sizes the ideogram and data plots and font sizes are automatically adjusted based on the length of the protein. Configuration and data files are exported for rendering of the image using Circos, which is available in Perl programming language.

PDBCirclePlot web servide stores the resulting images using a job ID that is calculated as the md5 hash of the parameters used for generating it. The user can later retrieve the same image using this job ID, without having to re-specify the paramters or uploading any custom data. PDBCirclePlot images of the proteins in the Protein Data Bank \citep{berman00}, generated using the default options, can be retrieved using the PDB ID, e.g., \url{http://sacan.biomed.drexel.edu/pdbcircleplot/img/1crn}.


%
%

%
%

\bibliographystyle{natbib}
\bibliography{bibliography}

\end{document}